\documentclass[conference]{IEEEtran}
\IEEEoverridecommandlockouts
\usepackage{cite}
\usepackage{amsmath,amssymb,amsfonts}
\usepackage{algorithmic}
\usepackage{graphicx}
\usepackage{textcomp}
\usepackage{xcolor}
\usepackage{booktabs}
\usepackage{url}
\usepackage{breakurl}
\usepackage{graphicx}

\usepackage{graphicx}   
\usepackage{float}      
\usepackage{tikz}
\usetikzlibrary{positioning, arrows.meta}
\usetikzlibrary{calc}

\def\BibTeX{{\rm B\kern-.05em{\sc i\kern-.025em b}\kern-.08em
    T\kern-.1667em\lower.7ex\hbox{E}\kern-.125emX}}
\begin{document}

\title{823-OLT @ BUET DL Sprint 4.0: Context-Aware Windowing for ASR and Fine-Tuned Speaker Diarization in Bengali Long Form Audio\\
{\footnotesize \textsuperscript{}}
\thanks{}
}

\author{
\IEEEauthorblockN{1\textsuperscript{st} Ratnajit Dhar}
\IEEEauthorblockA{
\textit{Department of Computer Science and Engineering} \\
\textit{Chittagong University of Engineering and Technology} \\
Bangladesh \\
u2004008@student.cuet.ac.bd}
\and
\IEEEauthorblockN{2\textsuperscript{nd} Arpita Mallik}
\IEEEauthorblockA{
\textit{Department of Computer Science and Engineering} \\
\textit{Chittagong University of Engineering and Technology} \\
Bangladesh \\
u2004023@student.cuet.ac.bd}
}

\maketitle

\begin{abstract}
Bengali, despite being one of the most widely spoken languages globally, remains underrepresented in long-form speech technology, particularly in systems addressing transcription and speaker attribution. We present frameworks for long-form Bengali speech intelligence that address automatic speech recognition using a Whisper-Medium–based model and speaker diarization using a fine-tuned segmentation model. The ASR pipeline incorporates vocal separation, voice activity detection, and a gap-aware windowing strategy to construct context-preserving segments for stable decoding. For diarization, a pretrained speaker segmentation model is fine-tuned on the official competition dataset (provided as part of the DL Sprint 4.0 competition organized under BUET CSE Fest), to better capture Bengali conversational patterns. The resulting systems deliver both efficient transcription of long-form audio and speaker-aware transcription to provide scalable speech technology solutions for low-resource languages.
\end{abstract}


\section{Introduction}
Long-form Bengali speech recognition (ASR) focuses on converting extended Bengali audio into accurate textual transcripts, while speaker diarization identifies “who spoke when” by segmenting audio according to speaker identity. These two components are fundamental for building practical speech intelligence systems that can handle multi-speaker conversational recordings.

Processing long-duration Bengali speech is particularly challenging due to background noise, music interference, irregular pauses, and speaker variability, especially in low-resource settings where large annotated datasets are scarce. Accurate ASR enables faithful speech-to-text conversion, while diarization preserves conversational structure by separating speaker turns in multi-party dialogue.

These technologies are especially valuable for Bengali, a language spoken by over 230 million people but still lacking sufficient speech resources compared to high-resource languages. Accurate long-form speech processing enables automated transcription of lectures, interviews, and meetings, and supports downstream tasks such as summarization, sentiment analysis, and speaker-level analytics in education, media, and customer service applications.

This work is developed as part of the DL Sprint 4.0 competition organized under BUET CSE Fest, 2026. The primary focus of this work is to develop a powerful and reproducible framework for long-form Bengali automatic speech recognition (ASR) and speaker diarization under realistic acoustic conditions, including multi-speaker conversations, background noise, silence intervals, and music contamination. Instead of relying on naive fixed-length segmentation, we design a pipeline that preserves contextual continuity while maintaining computational efficiency and inference stability.

To achieve this, we employ a gap-aware windowed decoding strategy using a pretrained Whisper-Medium ASR model, enabling coherent transcription across extended audio recordings. The system further incorporates structured preprocessing and segmentation mechanisms to improve resilience to acoustic disturbances. The diarization module is fine-tuned on the competition dataset to enhance speaker boundary detection and conversational speech modeling. The entire pipeline is implemented to be fully reproducible within the Kaggle runtime environment.

The main contributions of this work are summarized as follows:
\begin{itemize}
   \item A gap-aware windowing strategy for long-form ASR that reduces mid-sentence truncation while preserving transcription continuity.
\item A preprocessing pipeline for long-duration Bengali speech in noisy and music-contaminated audio.
\item A fine-tuned speaker segmentation model for Bengali diarization that surpasses the current benchmark.
\end{itemize}

Further implementation details can be accessed via the GitHub repository:\footnote{\url{https://github.com/ratnajit-dhar/823_OLT-BUET-DL-Sprint-Notebooks.git}}

\section{Related Work}

\subsection{Automatic Speech Recognition}

Modern ASR systems are primarily built on three objective families: Connectionist Temporal Classification (CTC)~\cite{Salazar_2019}, attention-based encoder--decoder models~\cite{chorowski2015attentionbasedmodelsspeechrecognition}, and Transducers (RNN-T)~\cite{graves2012sequencetransductionrecurrentneural}. Transformer-based encoders significantly improved long-range sequence modeling, while hybrid architectures such as Conformer~\cite{gulati2020conformerconvolutionaugmentedtransformerspeech} and ContextNet~\cite{han2020contextnetimprovingconvolutionalneural} incorporate local convolutional modeling for improved efficiency and stability.

Self-supervised learning further advanced ASR by leveraging large unlabeled corpora. wav2vec~2.0~\cite{baevski2020wav2vec20frameworkselfsupervised}, HuBERT~\cite{hsu2021hubertselfsupervisedspeechrepresentation}, and WavLM~\cite{Chen_2022} demonstrated strong transferability, while Whisper~\cite{radford2022robustspeechrecognitionlargescale} scaled weak supervision to 680k hours of multilingual data, enabling consistent cross-domain performance.

\subsection{Long-Form ASR}

Long-form ASR introduces challenges not prominent in short-utterance benchmarks, including segmentation errors, context fragmentation, speaker turns, overlap, and cumulative decoding drift. Naïve fixed-length chunking often leads to mid-sentence truncation and inconsistent context propagation.

Streaming-capable architectures such as Emformer~\cite{shi2020emformerefficientmemorytransformer} demonstrate bounded-memory designs for long-context modeling. In practice, effective long-form systems rely on VAD-guided segmentation, overlap-aware chunking, and careful merging strategies to balance contextual continuity with computational feasibility. Long-form ASR models may suffer from a long-form deletion problem, where extended blank predictions occur due to training–test mismatch; recent work mitigates this by incorporating speaker-aware modeling during decoding without additional computational cost~\cite{arumugam2023improvedlongformspeechrecognition}.

\subsection{Bangla ASR Resources}

Bangla ASR research has historically relied on limited open corpora. OpenSLR SLR53~\cite{kjartansson-etal-sltu2018} provides approximately 229 hours of read speech and has been widely used for baseline acoustic modeling. 
The OOD-Speech benchmark introduced by Bengali.AI~\cite{rakib2023oodspeechlargebengalispeech} contains over 1,100 hours of multi-domain speech, including dramas, talk shows, sermons, and online classes, specifically designed to evaluate performance under domain shift. Multilingual benchmarks such as FLEURS~\cite{conneau2022fleursfewshotlearningevaluation} also include Bengali for cross-lingual evaluation.

Despite these datasets, most Bangla corpora emphasize short or read speech. Long-form, multi-speaker recordings with diverse acoustic conditions remain relatively underrepresented.

\subsection{Speaker Diarization}

Traditional speaker diarization follows a segmentation $\rightarrow$ embedding $\rightarrow$ clustering $\rightarrow$ resegmentation pipeline. Speaker representations evolved from i-vectors~\cite{dehak2011ivector} to x-vectors~\cite{snyder2018xvectors} and ECAPA-TDNN~\cite{desplanques2020ecapa}, while clustering is often enhanced by probabilistic refinement such as VBx~\cite{landini2020bayesianhmmclusteringxvector}.

Long-form diarization exacerbates issues of global speaker consistency, re-entry, overlap, and domain mismatch, as highlighted in DIHARD evaluations~\cite{ryant2021diharddiarizationchallenge}. End-to-end approaches such as UIS-RNN~\cite{zhang2019fullysupervisedspeakerdiarization} and EEND~\cite{fujita2019endtoendneuralspeakerdiarization} attempt to jointly model speaker activity and overlap, with later variants supporting variable speaker counts~\cite{horiguchi2020eendeda}.

Recent benchmarks like Bengali-Loop introduced long-form datasets for Bangla ASR (158 hours) and diarization (22 hours), establishing baselines such as Tugstugi's Whisper variant at 34\% WER and pyannote at 40\% DER on multi-speaker content \cite{tabib2026bengali}.

\begin{figure*}[t]
\centering
\begin{tikzpicture}[
  node distance=6mm and 12mm,
  box/.style={
      draw,
      rounded corners,
      align=center,
      minimum height=7mm,
      minimum width=26mm,
      font=\footnotesize,
      fill=green!7
  },
  arrow/.style={-{Latex}, thick}
]

\node[box] (audio) {Input Audio};
\node[box, right=of audio] (demucs) {Demucs \\ (Vocal Separation)};
\node[box, right=of demucs] (prep) {Preprocessing \\ (Mono, 16kHz)};
\node[box, right=of prep] (vad) {Silero VAD};

\draw[arrow] (audio) -- (demucs);
\draw[arrow] (demucs) -- (prep);
\draw[arrow] (prep) -- (vad);

\node[box, below=14mm of audio] (win) {VAD-Aware \\ Windowing};
\node[box, right=of win] (pad) {Context Padding \\ ($\pm$1s)};
\node[box, right=of pad] (whisper) {Whisper-Medium \\ (Fine-tuned)};
\node[box, right=of whisper] (norm) {Text Normalization};
\node[box, right=of norm] (out) {Final Transcript};

\draw[arrow] (win) -- (pad);
\draw[arrow] (pad) -- (whisper);
\draw[arrow] (whisper) -- (norm);
\draw[arrow] (norm) -- (out);

\draw[arrow] (vad.south) -- ++(0,-4mm) -| (win.north);

\end{tikzpicture}
\caption{Long-form Bengali ASR System Architecture.}
\label{fig:asr_arch}
\end{figure*}
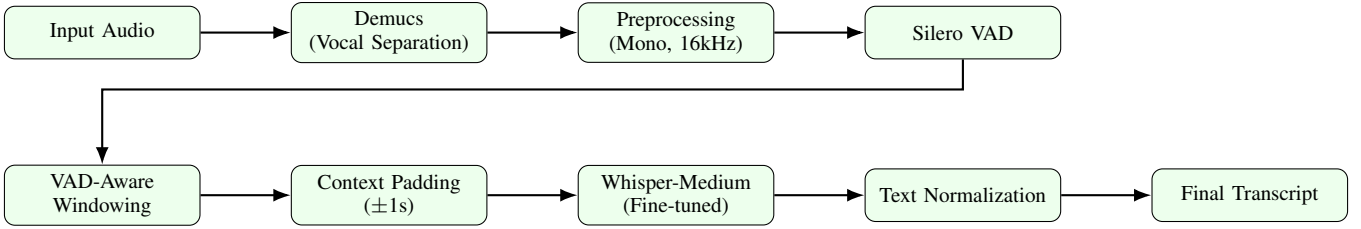

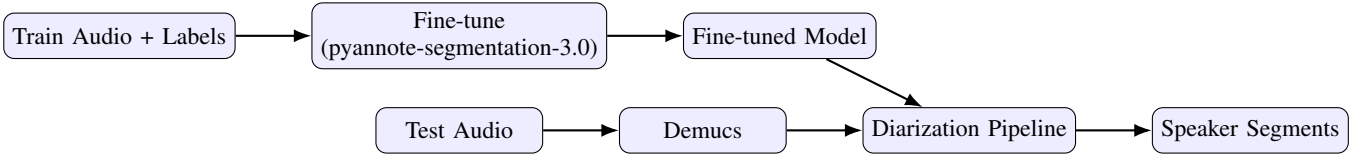
\begin{figure*}[t]
\centering
\resizebox{\textwidth}{!}{
\begin{tikzpicture}[
  node distance=6mm and 10mm,
  box/.style={
    draw,
    rounded corners,
    align=center,
    minimum height=6mm,
    minimum width=22mm,
    font=\small,
    fill=blue!7
  },
  arrow/.style={-{Latex}, thick}
]

\node[box] (train) {Train Audio + Labels};
\node[box, right=of train] (ft) {Fine-tune \\ (pyannote-segmentation-3.0)};
\node[box, right=of ft] (model) {Fine-tuned Model};

\node[box, below=5mm of ft] (test) {Test Audio};
\node[box, right=of test] (demucs) {Demucs};
\node[box, right=of demucs] (pipe) {Diarization Pipeline};
\node[box, right=of pipe] (out) {Speaker Segments};

\draw[arrow] (train) -- (ft);
\draw[arrow] (ft) -- (model);

\draw[arrow] (test) -- (demucs);
\draw[arrow] (demucs) -- (pipe);
\draw[arrow] (pipe) -- (out);

\draw[arrow] (model) -- (pipe);

\end{tikzpicture}
}
\caption{Diarization System Architecture.}
\label{fig:diar_arch_ft}
\end{figure*}

\section{Methodology}

\subsection{Long Form Automatic Speech Recognition}

Our long-form ASR system is designed as a modular pipeline to maintain stable transcription under noisy and multi-speaker conditions. The pipeline consists of: (1) vocal source separation, (2) audio preprocessing, (3) voice activity detection (VAD), (4) VAD-aware windowing, (5) Whisper-based decoding, and (6) text normalization.

\paragraph{Vocal Source Separation}
We apply Demucs (htdemucs)\footnote{\url{https://github.com/facebookresearch/demucs}} in two-stem mode to isolate vocals from background components. Only the extracted vocal track is used for ASR, improving transcription quality in recordings containing music or environmental noise. Experiments without Demucs resulted in higher transcription error rates, confirming the benefit of vocal enhancement in our pipeline.

\paragraph{Audio Preprocessing}
Audio is converted to mono and resampled to 16\,kHz. Let $x(t)$ denote the processed waveform used for downstream inference.

\paragraph{Voice Activity Detection}
Silero VAD~\cite{SileroVAD} is used to detect speech intervals 
\[
\{(s_i, e_i)\}_{i=1}^{n},
\]
where each segment corresponds to a speech region. If no speech is detected, an empty transcript is returned.

\paragraph{VAD-Aware Windowing}
To handle long recordings efficiently, speech segments are merged into windows of approximately 20 seconds. Consecutive segments are not merged if the silence gap exceeds 5 seconds. This preserves speech continuity while avoiding excessively long inputs.

\paragraph{Context Padding}
Each window $(t_s, t_e)$ is extended by 1 second of silence on both sides (within valid bounds) to reduce boundary artifacts during decoding.

\paragraph{ASR Model}
We use a fine-tuned Bengali Whisper-Medium model\footnote{\url{https://www.kaggle.com/datasets/tugstugi/bengali-ai-asr-submission}} via WhisperForConditionalGeneration. Each window is converted to log-Mel features using WhisperProcessor, followed by autoregressive decoding with a maximum of 256 generated tokens.

\paragraph{Text Normalization}
Final transcripts are normalized using Unicode NFC normalization, removal of zero-width characters, whitespace cleanup, and formatting standardization.

This design improves long-form transcription stability through speech-aware segmentation, noise suppression, and controlled decoding.

\subsection{Speaker Diarization}

We explored both training-free and supervised approaches for speaker diarization. Our development process progressed through three stages: (1) zero-training clustering-based pipelines, (2) off-the-shelf neural diarization using pyannote\cite{Bredin23}, and (3) fine-tuning of a neural segmentation model.

\subsubsection{Training-Free Diarization Pipelines}

As a baseline, we implemented a modular, training-free diarization pipeline. All variants shared the same front-end:

\textbf{Speech Enhancement:} Demucs two-stem vocal separation.
textbf{Speech Activity Detection:} Silero VAD.

\textbf{Windowing:} Fixed-length segmentation aligned with VAD boundaries.

\textbf{Speaker Embeddings:} ECAPA-TDNN embeddings\cite{desplanques2020ecapa} extracted for each speech segment.

Among the training-free approaches, spectral clustering with and without VBx refinement achieved the strongest performance, significantly outperforming DBSCAN and PLDA-based clustering. However, overall DER remained suboptimal for challenging long-form recordings.

\subsubsection{Off-the-Shelf Neural Diarization}

We next evaluated the end-to-end neural diarization pipeline provided by pyannote.audio\cite{Bredin23}. This approach leverages pretrained neural segmentation and clustering models trained on large-scale diarization datasets. 

Compared to purely clustering based pipelines, the pretrained neural approach showed improved stability in overlapping speech and varying acoustic conditions. Nevertheless, domain mismatch limited its performance on the competition data.

\subsubsection{Fine-Tuned Neural Segmentation Model}

To better adapt to the target domain, we fine-tuned the pyannote/segmentation-3.0 model on the provided training data using the huggingface diarizers library\cite{akesbi-diarizers}.

The best-performing training configuration is summarized in Table~\ref{tab:training_config}.

\begin{table}[t]
\centering
\caption{Best-Performing Training Configuration}
\label{tab:training_config}
\renewcommand{\arraystretch}{1.3}
\setlength{\tabcolsep}{6pt}
\begin{tabular}{|p{4cm}|p{3cm}|}
\hline
\textbf{Hyperparameter} & \textbf{Value} \\
\hline
Optimizer & AdamW \\
Learning Rate & $5 \times 10^{-4}$ \\
Batch Size & 8 \\
Gradient Accumulation & 2 steps \\
Number of Epochs & 12 \\
Learning Rate Scheduler & Cosine Decay \\
Warmup Ratio & 0.1 \\
Precision & Mixed Precision (FP16) \\
\hline
\end{tabular}
\end{table}

Multiple hyperparameter settings were evaluated during development; detailed comparisons are provided in Section~IV.

Our final diarization system integrates:

\begin{enumerate}
    \item Demucs-based vocal enhancement,
    \item Fine-tuned pyannote segmentation model for speech and speaker boundary detection,
    \item Neural clustering backend for speaker attribution.
\end{enumerate}

By combining speech enhancement with domain-adapted neural segmentation, the proposed system significantly reduced diarization errors compared to both training-free pipelines and off-the-shelf pretrained models.

Overall, fine-tuning the segmentation model proved critical for adapting to the acoustic and conversational characteristics of the competition dataset.

\section{Experimental Results and Comparisons}
This section summarizes the experimental outcomes of our systems for two Bengali long-form speech tasks: \textit{Automatic Speech Recognition (ASR)} and \textit{Speaker Diarization}. We report results under each task's official evaluation protocol and compare alternative modeling and preprocessing choices.

\subsection{Bengali Automatic Speech Recognition}
For the ASR task, performance is evaluated using Word Error Rate (WER). We report both Public and Private leaderboard scores, using the Private score as the primary indicator of generalization.

We study the impact of two components in the long form ASR pipeline: (i) Demucs based speech enhancement and (ii) VAD threshold tuning for segmentation prior to transcription, while keeping all other components fixed.

\begin{table}[t]
\centering
\caption{ASR results under different preprocessing configurations.}
\label{tab:asr_results}
\resizebox{1.01\columnwidth}{!}{
\begin{tabular}{lccc}
\hline
\textbf{Demucs} & \textbf{VAD Threshold} & \textbf{Private} & \textbf{Public} \\
\hline
Yes  & Default & 0.36240 & 0.34253 \\
Yes   & 0.3     & 0.36491 & 0.34112 \\
No   & 0.3     & 0.38920 & 0.37004 \\
Yes   & 0.4     & 0.36507 & 0.34123 \\
No   & 0.2     & 0.41841 & 0.37571 \\
\hline
\end{tabular}
}
\end{table}

It is evident that both speech enhancement and VAD configuration influence long-form ASR performance. Using Demucs with default VAD parameters gave the best Private score, suggesting modest benefits from preprocessing. The configurations with thresholds 0.3 and 0.4 show better and nearly identical performance in the Public scores.

In contrast, reducing the VAD threshold to 0.2 substantially degraded performance, likely due to increased false speech detections and noisier segments. The relative behavior remained consistent across Public and Private splits.

Overall, moderate VAD thresholds combined with optional speech enhancement yielded the most stable ASR performance in our experiments.

\subsection{Bengali Speaker Diarization}

Diarization performance was evaluated using Diarization Error Rate (DER). We report both Public and Private DER, using Private DER as the primary metric.

Among the explored learning rates, $5\times 10^{-4}$ yielded the most competitive results; therefore, only those configurations are presented in Table~\ref{tab:ft_5e4}. Evaluation of training-free pipelines are shown in Table~\ref{tab:zerotrain}.

\begin{table}[t]
\centering
\caption{Fine-tuned diarization results with learning rate $5\times 10^{-4}$ and warmup ratio $0.10$.}
\label{tab:ft_5e4}
\resizebox{1.01\columnwidth}{!}{
\begin{tabular}{ccccc}
\hline
\textbf{Epochs} & \textbf{Batch} & \textbf{Demucs} & \textbf{DER (Public / Private)} \\
\hline
12 & 4 & Yes   & 0.20249 / 0.29430 \\
12 & 8 & Yes   & \textbf{0.19284} / \textbf{0.29085} \\
15 & 8 & Yes   & 0.18758 / 0.29718 \\
18 & 8 & Yes   & 0.20406 / 0.28816 \\
18 & 8 & No   & 0.20292 / 0.28857 \\
\hline
\end{tabular}
}
\end{table}

\begin{table}[t]
\centering
\caption{Training-free Speaker Diarization Pipelines Scores.}
\label{tab:zerotrain}
\begin{tabular}{p{0.50\linewidth}cc}
\toprule
\textbf{Clustering / Refinement Variant} & \textbf{Private DER} & \textbf{Public DER} \\
\midrule
PLDA scoring $\rightarrow$ spectral clustering
& 0.53437 & 0.55581 \\

Spectral clustering + VBx refinement
& 0.42715 & 0.33578 \\

DBSCAN clustering
& 0.67882 & 0.69779 \\

Spectral clustering (no VBx)
& 0.42535 & 0.28062 \\

Pyannote without Training
& 0.44885 & 0.40084 \\
\bottomrule
\end{tabular}
\end{table}

From Tables~\ref{tab:ft_5e4} and~\ref{tab:zerotrain}, it is evident that Fine-tuned models consistently outperformed training-free variants. Although spectral clustering without refinement performs best among zero-training approaches, all such pipelines yielded substantially higher DER, indicating the importance of task-specific fine-tuning for long-form Bengali speech.

\section{Limitations}

While the proposed framework demonstrates strong performance on long-form Bengali audio, several limitations remain. The ASR component relies on a pretrained Bengali Whisper-Medium model without extensive domain-specific fine-tuning, which may reduce generalization to highly informal speech, strong regional accents, and unseen recording conditions. A key limitation of the system is its sensitivity to acoustically modified or edited speech; effects such as distortion, delay, reverberation, pitch shifting, and heavy noise degrade both transcription accuracy and speaker discrimination by introducing variability not fully captured during pretraining.

Although gap-aware windowing mitigates mid-sentence truncation, segmented decoding can still cause minor contextual inconsistencies. The diarization model may struggle with heavy speaker overlap, very short turns, highly similar vocal characteristics, or artificially altered voices that create intra-speaker variability. Finally, the full pipeline; including speech enhancement, neural segmentation, and autoregressive decoding, remains computationally demanding, limiting scalability in resource-constrained or real-time settings.

\section{Conclusion}

In this work, we presented a framework for long form Bengali speech processing where Automatic Speech Recognition and speaker diarization are treated as complementary but independently optimized components. The ASR system uses speech enhancement and VAD aware windowed decoding to maintain stable transcription over extended recordings, while the diarization module applies fine tuned neural segmentation to improve speaker boundary detection and attribution. Together, these components enable effective processing of long multi speaker recordings containing silence and background interference.

Future work will focus on improving tolerance to acoustically altered speech through augmentation based training, effect invariant modeling, enhanced preprocessing, and tighter integration between ASR and diarization for more coherent speech processing.

\section*{Ethics Statement}
This work strictly follows the competition’s rules on licensing, reproducibility, and data usage. All pretrained models are publicly available and properly licensed, no external APIs were used, and all training and inference were conducted within Kaggle as required. Any external data complied with the stated public data policy and did not involve the test set.

We also acknowledge privacy concerns and potential bias across accents and speaker groups, and emphasize responsible and fair use of speech technologies.

\bibliographystyle{IEEEtran} 
\bibliography{bibliography}    

@misc{akesbi-diarizers,
  author = {Kamil Akesbi and Sanchit Gandhi},
  title = {"Diarizers: A repository for fine-tuning speaker diarization models"},
  year = {2024},
  publisher = {GitHub},
  journal = {GitHub repository},
  howpublished = {\url{https://github.com/huggingface/diarizers}}
}

@inproceedings{Bredin23,
  author={Hervé Bredin},
  title={{pyannote.audio 2.1 speaker diarization pipeline: principle, benchmark, and recipe}},
  year=2023,
  booktitle={Proc. INTERSPEECH 2023},
}

@inproceedings{desplanques2020ecapa,
  title={{ECAPA-TDNN: Emphasized Channel Attention, propagation and aggregation in TDNN based speaker verification}},
  author={Desplanques, Brecht and Thienpondt, Jenthe and Demuynck, Kris},
  booktitle={Interspeech 2020},
  pages={3830--3834},
  year={2020}
}

@misc{SileroVAD,
  author = {Silero Team},
  title = {Silero VAD: pre-trained enterprise-grade Voice Activity Detector (VAD), Number Detector and Language Classifier},
  year = {2024},
  publisher = {GitHub},
  journal = {GitHub repository},
  howpublished = {\url{https://github.com/snakers4/silero-vad}},
}

@article{tabib2026bengali,
  title={Bengali-Loop: Community Benchmarks for Long-Form Bangla ASR and Speaker Diarization},
  author={Tabib, HM and Rifti, Istiak Ahmmed and Ehsan, Abdullah Muhammed Amimul and Dasgupta, Somik and Sowdha, Md Zim Mim Siddiqee and Sarker, Abrar Jahin and Nijamy, Md Rafiul Islam and Hossain, Tanvir and Khatun, Mst and Mahmood, Munzer and others},
  journal={arXiv preprint arXiv:2602.14291},
  year={2026}
  }

@inproceedings{Salazar_2019,
   title={Self-attention Networks for Connectionist Temporal Classification in Speech Recognition},
   url={http://dx.doi.org/10.1109/ICASSP.2019.8682539},
   DOI={10.1109/icassp.2019.8682539},
   booktitle={ICASSP 2019 - 2019 IEEE International Conference on Acoustics, Speech and Signal Processing (ICASSP)},
   publisher={IEEE},
   author={Salazar, Julian and Kirchhoff, Katrin and Huang, Zhiheng},
   year={2019},
   month=may, pages={7115–7119} }

@misc{chorowski2015attentionbasedmodelsspeechrecognition,
      title={Attention-Based Models for Speech Recognition}, 
      author={Jan Chorowski and Dzmitry Bahdanau and Dmitriy Serdyuk and Kyunghyun Cho and Yoshua Bengio},
      year={2015},
      eprint={1506.07503},
      archivePrefix={arXiv},
      primaryClass={cs.CL},
      url={https://arxiv.org/abs/1506.07503}, 
}

@misc{graves2012sequencetransductionrecurrentneural,
      title={Sequence Transduction with Recurrent Neural Networks}, 
      author={Alex Graves},
      year={2012},
      eprint={1211.3711},
      archivePrefix={arXiv},
      primaryClass={cs.NE},
      url={https://arxiv.org/abs/1211.3711}, 
}

@misc{gulati2020conformerconvolutionaugmentedtransformerspeech,
      title={Conformer: Convolution-augmented Transformer for Speech Recognition}, 
      author={Anmol Gulati and James Qin and Chung-Cheng Chiu and Niki Parmar and Yu Zhang and Jiahui Yu and Wei Han and Shibo Wang and Zhengdong Zhang and Yonghui Wu and Ruoming Pang},
      year={2020},
      eprint={2005.08100},
      archivePrefix={arXiv},
      primaryClass={eess.AS},
      url={https://arxiv.org/abs/2005.08100}, 
}

@misc{han2020contextnetimprovingconvolutionalneural,
      title={ContextNet: Improving Convolutional Neural Networks for Automatic Speech Recognition with Global Context}, 
      author={Wei Han and Zhengdong Zhang and Yu Zhang and Jiahui Yu and Chung-Cheng Chiu and James Qin and Anmol Gulati and Ruoming Pang and Yonghui Wu},
      year={2020},
      eprint={2005.03191},
      archivePrefix={arXiv},
      primaryClass={eess.AS},
      url={https://arxiv.org/abs/2005.03191}, 
}

@misc{baevski2020wav2vec20frameworkselfsupervised,
      title={wav2vec 2.0: A Framework for Self-Supervised Learning of Speech Representations}, 
      author={Alexei Baevski and Henry Zhou and Abdelrahman Mohamed and Michael Auli},
      year={2020},
      eprint={2006.11477},
      archivePrefix={arXiv},
      primaryClass={cs.CL},
      url={https://arxiv.org/abs/2006.11477}, 
}

@misc{hsu2021hubertselfsupervisedspeechrepresentation,
      title={HuBERT: Self-Supervised Speech Representation Learning by Masked Prediction of Hidden Units}, 
      author={Wei-Ning Hsu and Benjamin Bolte and Yao-Hung Hubert Tsai and Kushal Lakhotia and Ruslan Salakhutdinov and Abdelrahman Mohamed},
      year={2021},
      eprint={2106.07447},
      archivePrefix={arXiv},
      primaryClass={cs.CL},
      url={https://arxiv.org/abs/2106.07447}, 
}

@article{Chen_2022,
   title={WavLM: Large-Scale Self-Supervised Pre-Training for Full Stack Speech Processing},
   volume={16},
   ISSN={1941-0484},
   url={http://dx.doi.org/10.1109/JSTSP.2022.3188113},
   DOI={10.1109/jstsp.2022.3188113},
   number={6},
   journal={IEEE Journal of Selected Topics in Signal Processing},
   publisher={Institute of Electrical and Electronics Engineers (IEEE)},
   author={Chen, Sanyuan and Wang, Chengyi and Chen, Zhengyang and Wu, Yu and Liu, Shujie and Chen, Zhuo and Li, Jinyu and Kanda, Naoyuki and Yoshioka, Takuya and Xiao, Xiong and Wu, Jian and Zhou, Long and Ren, Shuo and Qian, Yanmin and Qian, Yao and Wu, Jian and Zeng, Michael and Yu, Xiangzhan and Wei, Furu},
   year={2022},
   month=oct, pages={1505–1518} }

@misc{radford2022robustspeechrecognitionlargescale,
      title={Robust Speech Recognition via Large-Scale Weak Supervision}, 
      author={Alec Radford and Jong Wook Kim and Tao Xu and Greg Brockman and Christine McLeavey and Ilya Sutskever},
      year={2022},
      eprint={2212.04356},
      archivePrefix={arXiv},
      primaryClass={eess.AS},
      url={https://arxiv.org/abs/2212.04356}, 
}

@misc{shi2020emformerefficientmemorytransformer,
      title={Emformer: Efficient Memory Transformer Based Acoustic Model For Low Latency Streaming Speech Recognition}, 
      author={Yangyang Shi and Yongqiang Wang and Chunyang Wu and Ching-Feng Yeh and Julian Chan and Frank Zhang and Duc Le and Mike Seltzer},
      year={2020},
      eprint={2010.10759},
      archivePrefix={arXiv},
      primaryClass={cs.SD},
      url={https://arxiv.org/abs/2010.10759}, 
}

@misc{arumugam2023improvedlongformspeechrecognition,
      title={Improved Long-Form Speech Recognition by Jointly Modeling the Primary and Non-primary Speakers}, 
      author={Guru Prakash Arumugam and Shuo-yiin Chang and Tara N. Sainath and Rohit Prabhavalkar and Quan Wang and Shaan Bijwadia},
      year={2023},
      eprint={2312.11123},
      archivePrefix={arXiv},
      primaryClass={cs.SD},
      url={https://arxiv.org/abs/2312.11123}, 
}

@inproceedings{kjartansson-etal-sltu2018,
    title = {{Crowd-Sourced Speech Corpora for Javanese, Sundanese,  Sinhala, Nepali, and Bangladeshi Bengali}},
    author = {Oddur Kjartansson and Supheakmungkol Sarin and Knot Pipatsrisawat and Martin Jansche and Linne Ha},
    booktitle = {Proc. The 6th Intl. Workshop on Spoken Language Technologies for Under-Resourced Languages (SLTU)},
    year  = {2018},
    address = {Gurugram, India},
    month = aug,
    pages = {52--55},
    URL   = {http://dx.doi.org/10.21437/SLTU.2018-11}
  }

@misc{rakib2023oodspeechlargebengalispeech,
      title={OOD-Speech: A Large Bengali Speech Recognition Dataset for Out-of-Distribution Benchmarking}, 
      author={Fazle Rabbi Rakib and Souhardya Saha Dip and Samiul Alam and Nazia Tasnim and Md. Istiak Hossain Shihab and Md. Nazmuddoha Ansary and Syed Mobassir Hossen and Marsia Haque Meghla and Mamunur Mamun and Farig Sadeque and Sayma Sultana Chowdhury and Tahsin Reasat and Asif Sushmit and Ahmed Imtiaz Humayun},
      year={2023},
      eprint={2305.09688},
      archivePrefix={arXiv},
      primaryClass={eess.AS},
      url={https://arxiv.org/abs/2305.09688}, 
}

@misc{conneau2022fleursfewshotlearningevaluation,
      title={FLEURS: Few-shot Learning Evaluation of Universal Representations of Speech}, 
      author={Alexis Conneau and Min Ma and Simran Khanuja and Yu Zhang and Vera Axelrod and Siddharth Dalmia and Jason Riesa and Clara Rivera and Ankur Bapna},
      year={2022},
      eprint={2205.12446},
      archivePrefix={arXiv},
      primaryClass={cs.CL},
      url={https://arxiv.org/abs/2205.12446}, 
}

@inproceedings{dehak2011ivector,
  author    = {Najim Dehak and Pedro A. Torres-Carrasquillo and Douglas A. Reynolds and Reda Dehak},
  title     = {Language Recognition via i-Vectors and Dimensionality Reduction},
  booktitle = {Proceedings of Interspeech 2011},
  pages     = {857--860},
  year      = {2011},
  doi       = {10.21437/Interspeech.2011-328}
}

@INPROCEEDINGS{snyder2018xvectors,
  author={Snyder, David and Garcia-Romero, Daniel and Sell, Gregory and Povey, Daniel and Khudanpur, Sanjeev},
  booktitle={2018 IEEE International Conference on Acoustics, Speech and Signal Processing (ICASSP)}, 
  title={X-Vectors: Robust DNN Embeddings for Speaker Recognition}, 
  year={2018},
  volume={},
  number={},
  pages={5329-5333},
  doi={10.1109/ICASSP.2018.8461375}}

@misc{landini2020bayesianhmmclusteringxvector,
      title={Bayesian HMM clustering of x-vector sequences (VBx) in speaker diarization: theory, implementation and analysis on standard tasks}, 
      author={Federico Landini and Ján Profant and Mireia Diez and Lukáš Burget},
      year={2020},
      eprint={2012.14952},
      archivePrefix={arXiv},
      primaryClass={eess.AS},
      url={https://arxiv.org/abs/2012.14952}, 
}

@misc{ryant2021diharddiarizationchallenge,
      title={The Third DIHARD Diarization Challenge}, 
      author={Neville Ryant and Prachi Singh and Venkat Krishnamohan and Rajat Varma and Kenneth Church and Christopher Cieri and Jun Du and Sriram Ganapathy and Mark Liberman},
      year={2021},
      eprint={2012.01477},
      archivePrefix={arXiv},
      primaryClass={eess.AS},
      url={https://arxiv.org/abs/2012.01477}, 
}

@misc{zhang2019fullysupervisedspeakerdiarization,
      title={Fully Supervised Speaker Diarization}, 
      author={Aonan Zhang and Quan Wang and Zhenyao Zhu and John Paisley and Chong Wang},
      year={2019},
      eprint={1810.04719},
      archivePrefix={arXiv},
      primaryClass={eess.AS},
      url={https://arxiv.org/abs/1810.04719}, 
}

@misc{fujita2019endtoendneuralspeakerdiarization,
      title={End-to-End Neural Speaker Diarization with Self-attention}, 
      author={Yusuke Fujita and Naoyuki Kanda and Shota Horiguchi and Yawen Xue and Kenji Nagamatsu and Shinji Watanabe},
      year={2019},
      eprint={1909.06247},
      archivePrefix={arXiv},
      primaryClass={eess.AS},
      url={https://arxiv.org/abs/1909.06247}, 
}

@article{horiguchi2020eendeda,
title = "Encoder-Decoder Based Attractors for End-to-End Neural Diarization",
author = "Shota Horiguchi and Yusuke Fujita and Shinji Watanabe and Yawen Xue and Paola Garcia",
note = "Publisher Copyright: {\textcopyright} 2014 IEEE.",
year = "2022",
doi = "10.1109/TASLP.2022.3162080",
language = "English",
volume = "30",
pages = "1493--1507",
journal = "IEEE/ACM Transactions on Audio Speech and Language Processing",
issn = "2329-9290",
publisher = "Institute of Electrical and Electronics Engineers Inc.",
}








%

\end{document}